# MR-compatible loading device for assessment of heel pad internal tissue displacements under shearing load.

**Authors:**

**Alessio Trebbi**
Univ. Grenoble Alpes, CNRS, TIMC-IMAG, 38000 Grenoble, France.
Alessio.Trebbi@univ-grenoble-alpes.fr

**Antoine Perrier**
Univ. Grenoble Alpes, CNRS, TIMC-IMAG, 38000 Grenoble, France
Groupe hospitalier Diaconesses–Croix Saint-Simon, 75020 Paris, France
TwInsight, 38000 Grenoble, France
perrier.antoine@gmail.com

**Mathieu Bailet**
TwInsight, 38000 Grenoble, France
mathieu.bailet@twinsight-medical.com

**Yohan Payan**
Univ. Grenoble Alpes, CNRS, TIMC-IMAG, 38000 Grenoble, France
Corresponding author. TIMC-IMAG Laboratory, Faculté de Médecine, Pavillon Taillefer, 38706 La Tronche cedex, France.
Yohan.Payan@univ-grenoble-alpes.fr

**Abstract**

In the last decade, the role of shearing loads has been increasingly suspected to play a determinant impact in the formation of deep pressure ulcers. In vivo observations of such deformations are complex to obtain. Previous studies only provide global measurements of such deformations without getting the quantitative values of the loads that generate these deformations. To study the role that shearing loads have in the aetiology of heel pressure ulcers, an MR-compatible device for the application of shearing and normal loads was designed. Magnetic resonance imaging is a key feature that allows to monitor deformations of soft tissues after loading in a non-invasive way. Measuring applied forces in an MR-environment is challenging due to the impossibility to use magnetic materials. In our device, forces are applied through the compression of springs made in polylactide. Shearing and normal loads were applied on the plantar skin of the human heel by means of an indenting plate while acquiring MR images. The device materials did not introduce any imaging artifact and allowed for high quality MR measurements permitting to identify the deformation of the internal components of the heel. The obtained subject-specific results are an original data set that can be used in validations for Finite Element analysis and therefore contribute to a better understanding of the factors involved in pressure ulcer development.

## 1. Introduction

Pressure ulcers are defined as localized areas of damaged skin and underlying soft tissues caused by sustained mechanical loads. They are especially common in patients who are bedridden, using a prosthesis or orthosis, or a wheelchair as these configurations generate continuous loadings in the interface between the body and the supporting surface. Ulcers are painful, hard to treat and represent a serious issue in terms of health care and money. As a result, they can affect the quality of life of both elderly and young individuals [1][2]. The clinical measures to prevent these types of injuries are challenged by the fact that they can occur inside the soft tissues (in case of deep tissue injury) and cannot in that case be detected by scanning the superficial layer of the skin. Additionally, it is still not completely understood how external loads are transmitted to local internal strains and, more importantly, how these lead to tissue damage. Many researchers have implemented Finite Element (FE) analysis to understand how external loads, applied on the skin surface, lead to deep internal strains [3][4]. Such an analysis is a powerful tool that permits to test many different configurations and obtain faithful results in terms of strain propagation [5][6]. However, validation of the corresponding FE simulations has always been problematic due to the difficulty to implement internal markers to track internal displacements of human soft tissues. A common technique to evaluate the predictions provided by FE models of the foot is to compare the simulated plantar pressures with measurements from pressure mats [7]. This solution offers the possibility to directly measure plantar pressures but lacks any analysis of the mechanical behaviors in the deeper tissues.

Until recently, prevention of pressure ulcers focused mainly on pressure distribution and little was made in order to analyze the harmful effects of shear-force, as pathophysiological aspects of shear-force are still not well known [8]. Only in the last decade, attention has been drawn on the role that shear strains could play in the development of deep tissue injuries [9][10].

Despite being the second most common type of pressure ulcers, heel ulcers are considerably understudied compared to the rest of the ulcer literature, having a relative low number of groups actively investigating in this area [6][11][12]. Most loadings applied to the human heel are localized in two regions, namely the posterior and the plantar regions of the heel. The plantar region is subject to repetitive normal and shearing loads generated by gait or standing positions [13][14]. It can also be subject to a constant load generated by the weight of the leg when a patient is sitting on a wheelchair [15]. On the other hand, the posterior region of the heel is subject to constant normal and shearing loads generated by the contacts and friction with the mattress in patients that are bedridden [6]. Despite the higher load intensity that is applied on the plantar region, ulceration has a much higher incidence on the posterior part of the heel [16][17]. This is due to a fatty layer, called fat pad, which covers the lower part of the heel. This structure is able to absorb impacts and redistribute the external loads to the internal tissues in order to prevent internal damage during gait. The fat pad is composed of adipose tissue for shock absorption and of a stiffer connective tissue (the septa) to maintain the pad structural integrity [18]. Currently, very few researchers are investigating the protective properties of the fat pad and analyzing the different roles of the materials it is made of [12].

Measuring *in vivo* the deformations of deep tissue is a complex procedure. Magnetic Resonance (MR) Imaging can be a solution for such a measurement as soon as an image is successively collected first with soft tissue at rest and second with the tissue deformed by an external loading device. These two images, if they are transformed into a common coordinate system using a 3D image registration technique, can provide a quantitative measurement of deep tissue displacement [19][20]. The main challenge to collect the two MR images consists in designing a loading device that is fully non-magnetic and that is able to quantitatively measure the applied load. Indeed, forces cannot be measured by conventional load cells that would disturb the magnetic field of the MR machine. Solutions to overcome this problem have been various in the literature. Chatzistergos et al. [21] used the gravitational force of a known MR-compatible mass to generate a specific load on the forefoot. Other applications used an optical sandwich [22], or a strain gauge placed sufficiently far from the region of interest (ROI) to measure the applied load [23]. However, with these solutions, the applied forces were limited to 5 Newtons (N) which is considerably lower than the forces observed during gait [13]. Devices that were capable of applying higher loads used pistons actuated by hydrostatic pressure that was remotely controlled outside the MR room [24]. Among these solutions, the device proposed by Petre et al. [25] was the only one designed to apply and to measure shearing loads on the metatarsal region. However, to the best of our knowledge, such functionality of the device was never used in practice: in subsequent works, only the normal loads have been applied [26].

At this point, it is possible to list the requirements that an MR compatible device should have in order to address the matters described above. First of all, it is important to state that the device should provide a dataset (in terms of 3D images and boundary conditions) that will allow a complete FE analysis of heel pad internal tissue deformations. With this respect, the device should be able to apply loadings on the skin's surface (with a quantitative measurement of these loads) and to measure the generated internal tissue displacements. The type of applied loading should be both with normal and shearing forces since shear forces have been increasingly suspected to play a determinant impact in the formation of deep pressure ulcers [27]. Finally, the applied loading should be comparable with the one observed during gait. This paper aims at providing a response to such requirements, with the design of an MR compatible device able to apply calibrated shearing and normal loads and with the use of a non-rigid image-based registration method that provides a quantitative measurement of internal tissue displacements.

## 2. Methods
### 2.1. MR-compatible loading device

The loading device consists in two parts: the foot fixation and the mechanics of the loading plate. (Scheme of the device in Figure 1 and photos of that device in Figure 2). These two components are held together by an external frame made in polyvinyl chloride (PVC). The loading mechanics are based on the compression of two sets of polylactide (PLA) springs respectively oriented to generate a normal and shearing load. Springs were printed using conventional 3D printers with 100% infill. The printing orientation was set in order to have the spring axis orthogonal to the printing plane. Table 1 gives an overview of the springs implemented in the loading device.

|  | Configuration | Coil Diameter [mm] | Length [mm] | Spring Diameter [mm] | Nb. of coils |
|---|---|---|---|---|---|
| Normal load set | 4 parallel | 5.0 | 55.0 | 25.0 | 4 |
| Shear load set | 2 parallel | 3.0 | 43.0 | 20.0 | 4 |

*Table 1 Parameters of springs implemented in the loading device*

By rotating a handle connected to a nylon screw (in grey color on Figure 1) the operator is able to displace one extremity of the springs. The other extremity is directly connected to the indenting plate. By displacing the extremity connected to the nylon screw, the spring compresses and exerts a force on the plate that delivers the load on the heel. Figure 1 provides a comparison of the device for an unloaded configuration (Load 0) and a loaded configuration where the plate is applying a normal and a shearing load (Load N). The directions of the loads are shown in the figure from the green and orange arrows, respectively representing the shearing $F_s$ and normal $F_n$ forces.

To reduce frictions, the plates connected to the spring extremities are constrained by guides made in Teflon which respectively allow one degree of freedom. A two sided tape was placed on the indenting plate to bind the heel to that plate and to ensure no sliding while applying the shearing loads. The foot casing (in blue in Figure 1) was used to hold the heel in a fixed position by blocking the rotation of the ankle and limiting any additional displacement of the foot with respect to the MR machine. This component was designed with a negative of the surface of the foot obtained in a previous MR acquisition and manufactured in PLA with the use of a 3D printer (see figure 2 for a view of this foot casing). The casing was printed in order to allow 1 mm of margin with the foot while the sharp edges in contact with the skin were filleted in order to make the device more comfortable to wear. The external frame and foot casing were held together by screws made in nylon.

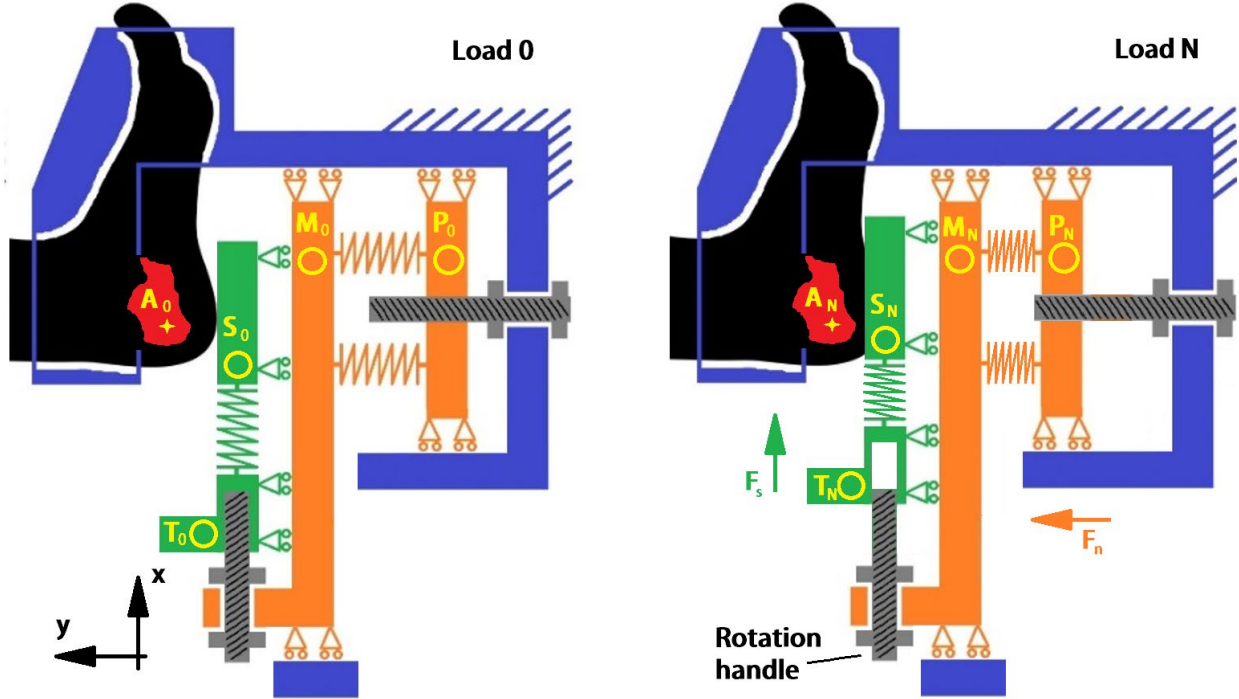

*Figure 1: Scheme of the MR-compatible compression device. The human heel (in black) is deformed by an indentation plate (in green) that exerts normal and shearing forces (resp. Fn and Fs when Load N is applied). External fixed frame is colored in blue. The mechanism to apply normal loading Fn is shown in orange while the mechanism to apply shearing loading Fs is shown in green. The handles and respective screws to displace the spring extremities are shown in grey. Calcaneus bone is drawn in red. MR markers are plotted as yellow circles. From the MR acquisition, only components colored in red, yellow and black are visible in the images. However, thanks to the MR markers, it is still possible to deduce the position of every component at the time of the acquisition. Coordinate frame is set in order to make more intuitive the subsequent analysis.*

MR markers, illustrated with yellow circles in Figure 1, were glued on the key components of the device to measure the level of compression of the springs along the acquisition. The image of the calcaneus, captured in the acquisition allows for the identification of key points (landmarks like small holes in the bone, illustrated as point A in Figure 1) that permit to identify the internal displacement of the calcaneus between two images. The position of the calcaneus together with the position of the plate allows to calculate the indentation of the fat pad soft tissues. Spring compressions and tissue indentations were calculated from the positions (in the x-y sagittal plane) of the MR markers with the following formulas:

$Spring\ normal\ deflection = (M_{0y} - P_{0y}) - (M_{Ny} - P_{Ny})$ (1)
$Spring\ shear\ deflection = (S_{0x} - T_{0x}) - (S_{Nx} - T_{Nx})$ (2)
$Soft\ tissue\ normal\ indentation = (A_{0y} - S_{0y}) - (A_{Ny} - S_{Ny})$ (3)
$Soft\ tissue\ shear\ indentation = (A_{0x} - S_{0x}) - (A_{Nx} - S_{Nx})$ (4)

Measuring for how long the springs were compressed during the experiment allows to reproduce *a posteriori* the compression time history of the springs during the experiment. The compression timings can then be reproduced later, in a compression test machine, by applying the same compressions captured in the MR images for the same time frame of the experiment. This gives an estimation of the forces that the device applied on the soft tissues across the MR acquisition.

## 2.2. Experimental protocol

A volunteer (male, 40 years old) agreed to participate in an experiment part of a pilot study approved by an ethical committee (MammoBio MAP-VS pilot study N°ID RCB 2012-A00340-43, IRMaGe platform, Univ. Grenoble Alpes). He gave his informed consent to the experimental procedure as required by the Helsinki declaration (1964) and the local Ethics Committee. The volunteer was placed in a supine position with his right foot inserted into the compression device and fixed inside the foot casing. A proton density MR was used to collect 512 consecutive 0.3125 mm thick sagittal slices (MRI system Achieva 3.0T dStream Philips Healthcare). Each slice had a field of view of 160 × 160 mm and a resolution of 512 × 428 for a total scanning time of 7 minutes.

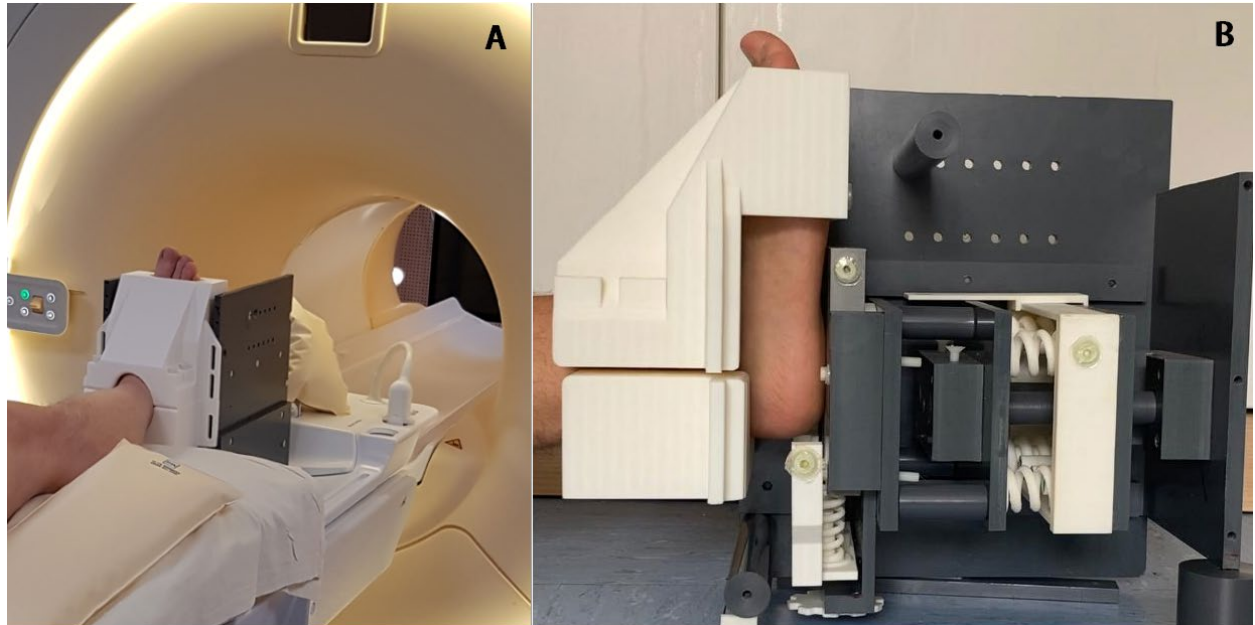

*Figure 2: A) Loading device attached to the foot of the subject entering the MR machine. The white part is the 3D printed foot casing used to hold the heel in a fixed position. B) Internal view of the MR compatible device showing the springs and their orientation. The white gear wheel that can be seen in the lower part of the image is the rotating handle through which the user can displace one extremity of the springs. The circles of around 2 cm in diameter attached on the components are the MR markers.*

The imaging acquisition protocol was divided into five stages in order to capture images for five different loading configurations. The first acquisition, Load 0, was taken in a resting configuration with the indenting plate in a tangent position with respect to the plantar skin of the heel but without any contact with the heel. This step was also used to identify the proper region of interest (ROI) and to make sure all the MR markers, together with the relevant part of the heel, were present in the 3D images. Then, four successive load configurations were applied by rotating the handles of the device to displace the internal springs:

- Load 0 – No load applied
- Load 1 – High normal load
- Load 2 – Mild normal load
- Load 3 – Mild normal load and high shearing load
- Load 4 – Mild normal load and mild shearing load

This was chosen in order to capture the non-linear mechanical properties of the soft tissues: stiffer for higher loads and more compliant to lower loads. The same milder normal force was kept through the application of the shearing loads in order to ensure good contact between the indenting plate and the plantar skin of the heel.

### 2.3. Force measurements

The MRI acquisitions were analyzed using Amira software. Image segmentations of the heel at rest were carried out in order to define regions corresponding to bone, muscle, tendon, fat and skin. In order to align all 3D images, a rigid registration (rotation + translation) was implemented using the publicly available registration package Elastix [28]. Images were aligned with respect to the position and orientation of the calcaneus. For this purpose, a mask hiding soft tissues around the calcaneus was implemented. This was done in order to concentrate the registration capabilities on the alignment of the calcaneus that is considered to be undeformed in all the loading configurations. The Load 0 configuration is settled as the "fixed image" and the other loads (Load 1-4) to be "moving images". This was chosen in order to get all the displacement fields starting from the same initial configuration (Load 0).

Once images were aligned, the locations of the MRI markers were considered. By comparing the respective positions of these markers in the images, the compressions of the springs were calculated for each loading configuration (Equations 1 and 2). In order to estimate the applied forces during the acquisitions, the spring compressions for the normal and shearing load were *a posteriori* reproduced in a compression test machine (MTS Criterion Model 41): the compression plates deflected the sets of springs for the same amount displayed in MR images and for the time frame noted during the session. The same compression test was repeated 8 times for each set of springs in order to analyze possible stiffness changes across subsequent compressions.

### 2.4. Internal tissue displacement

In order to measure the tissue internal displacements, a non-rigid image registration method using the Elastix package was implemented with the Normalized Correlation Coefficient [28] used as the similarity measure between images. The aligned set of images obtained by the previous rigid registration were implemented as inputs for this non-rigid registration in order to have the volume of the calcaneus aligned in the images. The Load 0 configuration was imposed to be the fixed image and the others (Load 1-4) to be moving ones (see [28] for a description of these notions of fixed and moving images in Elastix). This allowed to obtain the respective displacement fields for each loaded configuration.

## 3. Results

### 3.1. MRI acquisition

The developed MR-compatible loading device allowed high quality MR measurements of the heel before and during the application of loads by means of the indenting plate. A good fixation on the ankle and the metatarsal part of the foot allowed to maintain the heel in position for well-defined consecutive MR measurements and image comparisons. No image artifact was detected in the ROI (Figure 3). The MR acquisitions had enough quality to clearly distinguish the different tissues that compose the human heel: bone, muscle, tendons, fat and skin. This allowed for a semiautomatic segmentation of the respective tissues. In the following images the result of the segmentation of the heel at rest (in red on Figure 3) are superposed in order to help the comparison between the loading configurations. To analyze the effect of the shear loading, we will focus on Loads 2 and 3 as their conditions are the same, apart for the addition of the shearing component in Load 3. The superposition of the red continuous line with the calcaneus can be used to evaluate the result of the alignment process.

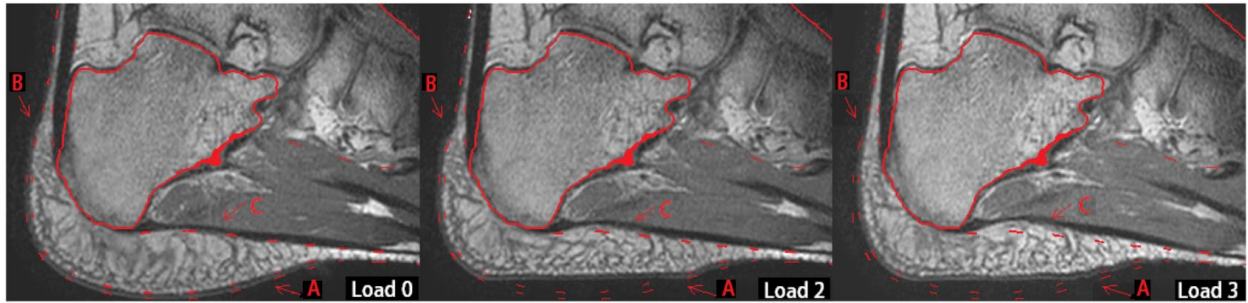

*Figure 3: Sagittal MRI sequence of loading configurations in order to compare the effects of normal and shearing loadings. The red contours represent the segmentation of Load 0. Arrows A, B and C are placed in locations where the most clear impacts of the applied loads are visible. Load 0: no load ; Load 2: Fn as a normal load ; Load 3: Fn + Fs as a normal and shearing load.*

The effects of the applied loads can be inspected visually in Figure 3. Arrow A points to the location where the plate was applying the loading. As can be seen for Load 2, the normal load dislocates the adipose tissue of the fat pad from the lower part of the calcaneus to the sides and the posterior part of the heel, as pointed by arrow B. Once the shearing load is added (Load 3), the fat is moved from the posterior part to the lower part of the calcaneus close to where the plantar fascia is connected to the calcaneus (arrow C).

### 3.2. Force measurement

The MR images allowed to identify the position of the MR markers to reconstruct the amount of compression of the springs and the position of the indenting plate during the experiment. Figure 4 illustrates heel deformations due to Load 1 with a superposition of the spring as represented in Figure 1. The components that were not captured in the MR acquisition as the indenting plate in green and the springs in red were added for clarification.

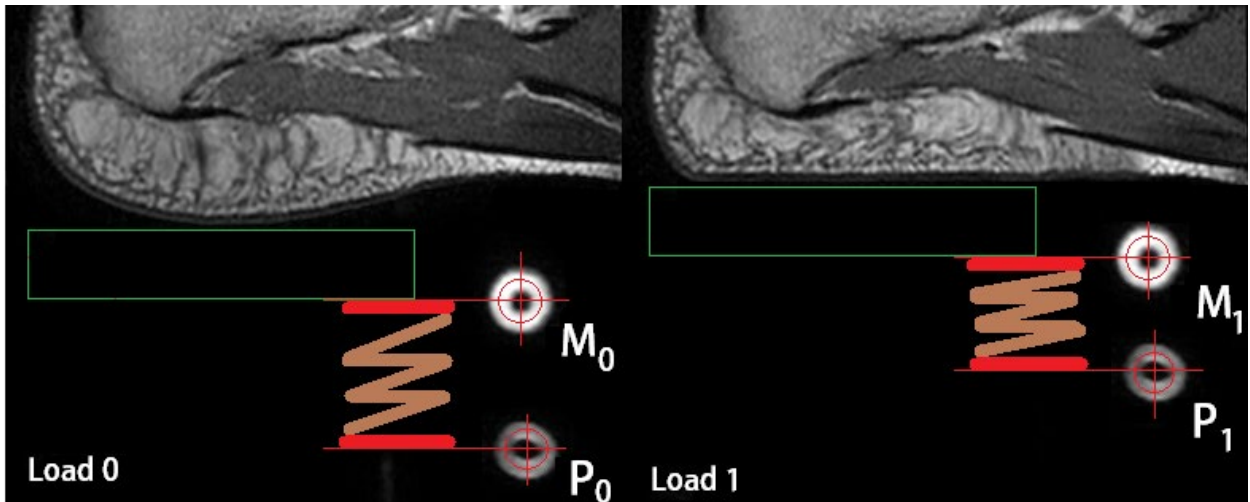

*Figure 4 Process of tracking the positions of the MR markers in order to calculate the level of compression of the springs. This figure is the actual result of Figure 1 for Load 1. The MR markers represented by yellow circles in Figure 1 are the only components of the device that can be seen in the images, together with the biological tissues. The two circles represent the spring extremities. In Load 0 the spring is in a relaxed position with the markers far apart. For load 1, the circles are close to each other showing that the spring was compressed. The upper marker is also closer to the foot sole, showing that the indenting plate moved upwards to compress the heel.*

Figure 5 plots the results of the four loading configurations measured from the springs' compressions and reproduced in the compression test machine. The 3D printed springs implemented in this study showed a viscoelastic behavior that reduced the reaction force of the spring over time. In order to reduce the force drop, the MR acquisitions were performed 3 minutes after compressing the springs. Areas highlighted in green show the time frames in which the machine was acquiring the images. The respective shaded areas in blue and red show the standard deviations (S.D.) across the 8 compression trials. Table 2 resumes key points for each loading configuration.

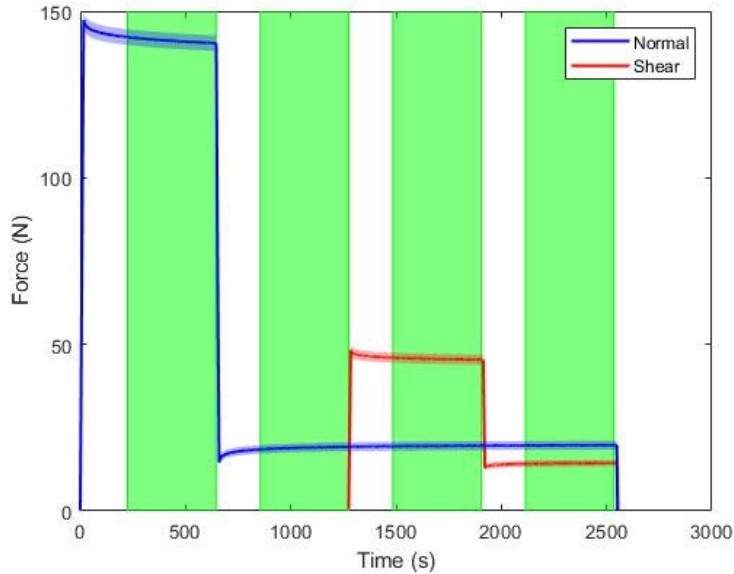

*Figure 5 A posteriori reproduction of springs loading configurations in the compression test machine. Green regions show when the MR was acquiring images. Load 1 is related to the first green area, Load 2 to the second and so on.*

|        | Normal force [N] | S.D. | Normal force drop [N] | Shear force [N] | S.D. | Drop [N] |
|--------|------------------|------|-----------------------|-----------------|------|----------|
| Load 1 | 140.35           | 2.34 | -1.63                 | 0               | 0    | 0        |
| Load 2 | 14.86            | 1.33 | +0.80                 | 0               | 0    | 0        |
| Load 3 | 15.85            | 1.33 | +0.20                 | 45.05           | 1.55 | -0.40    |
| Load 4 | 16.15            | 1.33 | +0.10                 | 11.87           | 1.00 | +0.23    |

*Table 2 Values resulting from the reproduction of springs loading configurations in the compression test machine. The force drop refers to the reduction in force due to the viscoelastic behaviour of the springs.*

### 3.3. Plate force/displacement pseudo-curves

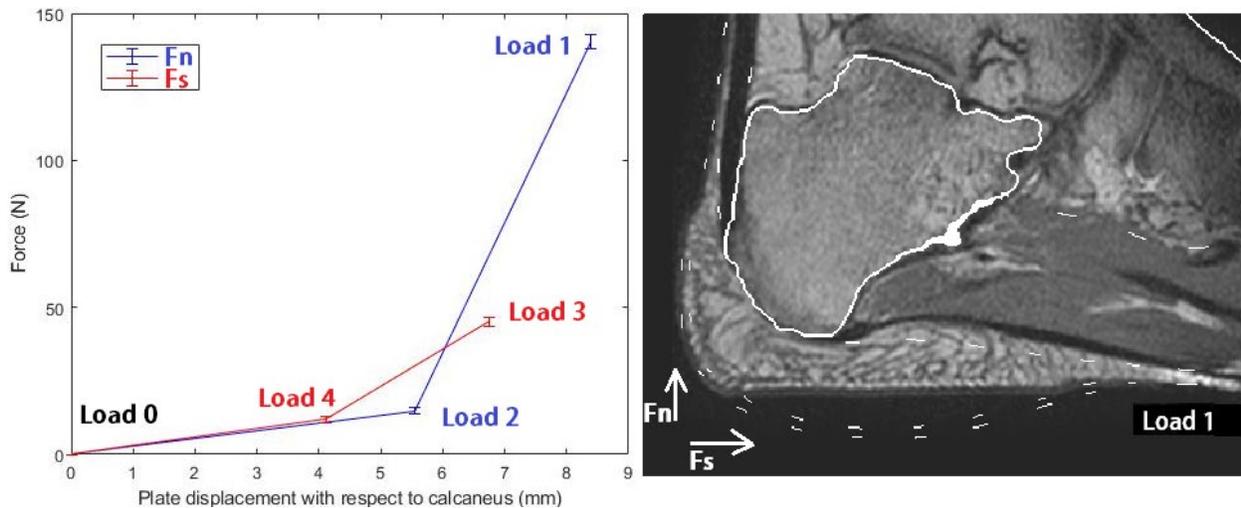

*Figure 6: A) Plots the loading configurations on a force/displacement plane. The blue points only refers to the normal forces applied by the normal set of springs and the respective displacement of the plate with respect to the calcaneus. The red points only refer to the shearing forces applied by the springs. B) Slice from the Load 1 configuration that had the highest normal load applied in the experiment: 140 N. The two arrows Fn and Fs show the direction of the appled loads.*

Figure 6 plots in 2D the four loading configurations applied during the experiment with the forces generated by the springs on the vertical axis and the displacement of the indenting plate with respect to the calcaneus on the horizontal axis. In this figure we want to distinguish the behavior of the soft tissues under normal load and shearing load. With this regard, the blue line of the graph is only related to load 1 and 2 and to the forces and displacements applied in the direction normal to the heel sole surface (i.e. the y direction as represented in Figure 1). On the other hand, the red line focuses on Load 3 and 4 that applied the shearing loads and only account for forces and displacements applied in the direction tangential to the heel sole surface (i.e. the x direction as represented in Figure 1). Values for the plate displacements were calculated using equations 3 and 4, while the force values are related to Figure 5. As can be seen in Figure 6, both the normal and shearing loadings reproduced a non-linear response of the plantar tissues: initial low stiffness followed by a higher stiffness when the applied force increases (which is coherent with what is known as concerns the non-linear constitutive behavior of most human soft tissue [7][29]).

### 3.4. Internal tissue displacements

This section aims at analyzing the displacements of fat pad internal tissues due to the applications of normal and shearing loads. The presented results are obtained from the non-rigid image registration performed between Load 0 and each other loading configurations (Load 1-4). Since the images were already aligned with respect to the calcaneus bone, the displacement of the points located inside that bone are expected to have a displacement close to zero. In order to compare tissue displacements due to normal loads versus displacements due to shearing loads, we show the respective slice of Load 2 and 3 previously presented in Figure 3.

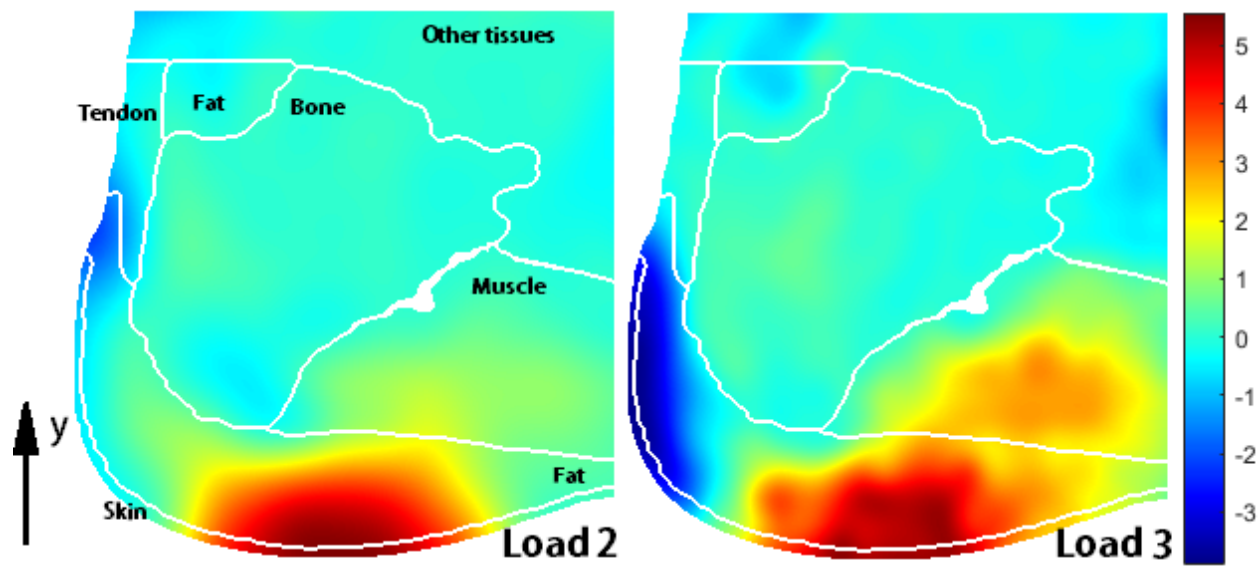

*Figure 7 Sagittal view of the internal tissues displacement map (in mm) obtained from the non-rigid image registration. Note that only the components of the displacements (in mm) along the direction normal to the heel sole surface (y axis) are plotted. White lines mark the boundaries of the biological tissues. Note that the displacement map is plotted in the heel geometrical configuration that corresponds to the undeformed state (as collected for Load 0). Load 2: mild normal load ; Load 3: mild normal load + high shearing load*

Figure 7 plots a sagittal view of the internal displacements (in mm) with respect to the calcaneus along the direction normal to the heel sole surface (i.e. the y direction as represented in Figure 1) for Loads 2 (left panel) and 3 (right panel). Results between Load 2 and 3 are similar in the lower part of the heel (arrow A in Figure 3) as the amount of normal load is the same, but some relevant differences appear when the shearing force is applied (Load 3). As previously noted by the arrow B in Figure 3, the shearing load is pulling downwards the posterior tissues of the heel. The effects of the shearing load can also be analyzed in the plantar aponeurosis region (Arrow C in Figure 3), where the tissues are pushed upwards.

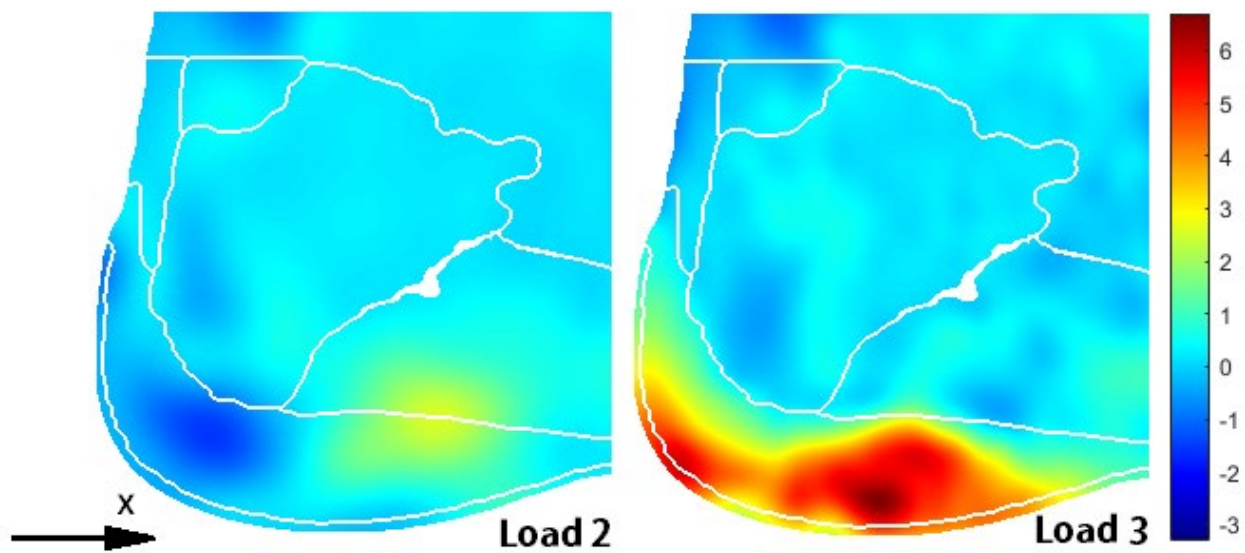

*Figure 8: Analogous plot to Figure 7 but, in this case, only the displacement component along the direction tangential to the heel sole surface (i.e. the x direction as represented in Figure 1) are plotted.*

Figure 8 is the analogous version of Figure 7 but with respect to the displacements along the direction tangential to the heel sole surface (i.e. the x axis). In this case, plots look different since no shearing force in the x direction is exerted by the plate for Load 2. For that load, the calcaneus is acting as a watershed pushing the fat towards the back or the front according to its initial location (the dark blue spot towards the left part of the bone surface and the yellow spot towards the right part). Displacements become significantly higher looking at Load 3, where the shearing load is clearly pushing all the tissues towards the frontal part of the foot.

## 4. Discussion

A MR-compatible loading device was developed to permit the simultaneous application of normal and shearing loadings on the plantar part of the human heel while acquiring MR images. A rigid and non-rigid image-based registration method was then used to align the images and to quantitatively measure the internal tissue displacements. The quality of the MR acquisitions and the absence of artifacts is enough for a semi-automatic segmentation process that defines surfaces delimiting the internal anatomical components of the heel: bone, muscle, tendon, fat and skin. The quality of the images allowed to successfully align the calcaneus bone surfaces of each image, thus providing a rigid registration of the different loading configurations between each other. This can be seen in Figure 3 where the calcaneus bone surface segmented from Load 0 perfectly fits the calcaneus frontiers as shown in the images related to Load 2 and 3.

The application of shearing and normal loads generated considerable displacements in the fat pad of the subject that can be visually identified in the MR images (Figure 3). The applied force was estimated considering the compression of 3D PLA springs. The force drop due to the viscoelastic properties of the PLA springs had a maximal impact in Load 1 configuration with a drop in force of 1.63 N over 140 N. Repeating the same compression test for the springs has shown not significant changes in stiffness of the 3D springs across subsequent usages. The highest standard deviation was found in Load 1 with 140.35 N ± 2.34 N (i.e. an accuracy of 1.6%). Information about force accuracy of other MR-compatible loading devices is scarce and is mostly presented in terms of pressure. In the works of Stekelenburg et al. [23] and Tokuno et al. [30], force accuracy is less than 1%. However, it must be considered that these devices are capable of measuring forces that do not reach values higher than 5 N and therefore are not suitable for applications related to the plantar pressures that have a consistent higher value. Thanks to the ease of 3D printing, in our device, various sets of springs with different parameters can be adapted to reach any required amount of stiffness and accuracy. The main limitation of this device is the impossibility to apply dynamic loadings and to directly control during the MR acquisition the amount of force applied. The level of force can indeed only be estimated *a posteriori* with an analysis of the MR images and a compression test of the relative springs. This limitation, however, is common in devices of this kind due to the limited allowed materials in the MR environment [23][25]. A possibility to overcome this limitation would be to *a priori* mechanically characterize the set of springs and to calibrate them in order to know the exact amount of force exerted by the plate for any compressions of the springs measured during the MR session. This would however not be a straightforward task as it would require an on-line segmentation of the MR images to track the MR markers or, as an alternative, it would require to design a mechanical system to accurately measure the springs compressions during the MR session (like for example a 3D printed caliber attached to the moving parts of the device). This could be explored as a perspective of this study.

Displacements computed from the Elastix non-rigid image registration are in line with what was expected looking at Figure 3. First, loads applied in a direction normal to the heel sole surface compress tissues against the calcaneus frontier and push some tissue at the back of the foot, in the region of the Achilles tendon. Second, shearing loads mainly moved the soft tissues towards the frontal part of the foot. It is also interesting to note that the higher displacement values at skin external surface as plotted for Figure 7 load 2 and Figure 8 load 3 match with the values reported in the plot of Figure 6 (and measured from the location of MR markers). This is reasonable as we expect that the displacement of the first layer of skin matches with the displacement of the indenting plate (as skin is fixed to the plate with two-sided tape).

## 5. Conclusion

This paper has proposed the design of a MR-compatible device capable of applying shearing and normal loadings on the human heel and of quantitatively measuring the corresponding force and internal displacement values. The device is based on the compression of 3D printed springs. The imaging process produced a valid data set for the analysis of displacements of the internal components of the fat pad. The outcomes of this study are multiple. First, the possibility to apply shearing loads and to observe the corresponding tissue deformations provide insights as concerns the impact that these loads can have on the biological tissues in the development of pressure ulcers. Second, the displacement of internal soft tissues generated by the high loadings is crucial for understanding the roles of the fat pad sub-components in mitigating the formation of pressure ulcers in the plantar part of the heel. Third, the load/displacement pseudo-curves can be used in the estimation of the biomechanical properties of the heel pad subcomponents, especially for FE models that require constitutive parameters. Finally, the data related to the deformations of internal soft tissues can be used to validate the simulations provided by such FE foot models to reproduce gait or configurations related to wheelchair or bedridden patients.

This work represents a pilot study aiming to investigate whether the crucial components of this procedure are feasible to be implemented with more subjects and different loading configurations. Further research will include analysis on more subjects and with different loading configurations, together with the adaptation of the device to apply loadings on different parts of the body to gain insight into the relative mechanical soft tissue properties.

**Acknowledgments**

This research has received funding from the European Union's Horizon 2020 research and innovation programme under the Marie Skłodowska-Curie Grant Agreement No. 811965; project STINTS (*Skin Tissue Integrity under Shear*). IRMaGe MRI facility was partly funded by the French program "Investissement d'Avenir" run by the "Agence Nationale pour la Recherche" ; grant "Infrastructure d'avenir en Biologie Sante" - ANR-11-INSB-0006.

Competing interests: None declared